\documentclass[10pt,conference]{IEEEtran} \IEEEoverridecommandlockouts
\usepackage{cite}
\usepackage{amsmath,amssymb,amsfonts}
\usepackage{algorithmic}
\usepackage{graphicx}
\usepackage{textcomp}
\usepackage{xcolor}

\usepackage{multirow}
\usepackage{tabularx}
\usepackage{booktabs} %
\usepackage{array}    %
\usepackage[margin=1in]{geometry} %
\usepackage{url}

\usepackage{authblk}

\newcommand{\code}[1]{\texttt{#1}}

\def\BibTeX{{\rm B\kern-.05em{\sc i\kern-.025em b}\kern-.08em
    T\kern-.1667em\lower.7ex\hbox{E}\kern-.125emX}}
\begin{document}

\title{Identifying Flaky Tests in Quantum Code: A Machine Learning Approach}

\author[]{Khushdeep Kaur\textsuperscript{*}\thanks{\textsuperscript{*}The first two coauthors have made equal contributions to this work.}}
\author[]{Dongchan Kim\textsuperscript{*}}
\author[]{Ainaz Jamshidi}
\author[]{Lei Zhang}

\affil[]{Department of Information Systems, University of Maryland, Baltimore County, MD, USA}

\affil[ ]{{\{khushdk1,dkim26,ainazj1,leizhang\}@umbc.edu}}

\maketitle

\begin{abstract}
Testing and debugging quantum software pose significant challenges due to the inherent complexities of quantum mechanics, such as superposition and entanglement. One challenge is indeterminacy, a fundamental characteristic of quantum systems, which increases the likelihood of flaky tests in quantum programs. To the best of our knowledge, there is a lack of comprehensive studies on quantum flakiness in the existing literature. In this paper, we present a novel machine learning platform that leverages multiple machine learning models to automatically detect flaky tests in quantum programs. Our evaluation shows that the extreme gradient boosting and decision tree-based models outperform other models (i.e., random forest, $k$-nearest neighbors, and support vector machine), achieving the highest F1 score and Matthews Correlation Coefficient in a balanced dataset and an imbalanced dataset, respectively. Furthermore, we expand the currently limited dataset for researchers interested in quantum flaky tests. In the future, we plan to explore the development of unsupervised learning techniques to detect and classify quantum flaky tests more effectively. These advancements aim to improve the reliability and robustness of quantum software testing.
\end{abstract}


\section{Introduction}\label{sec:Intro}
Flaky tests are tests that inconsistently pass or fail across multiple runs without any changes to the code. They pose a significant issue in software testing and lead to uncertainty because of indeterministic results. In one study, Memon et al. show that 46,694 out of 115,160 (i.e., about 41\%) tests that pass sometimes and fail sometimes are found to be related to flaky in Google’s Test Automation Platform~\cite{memon2017taming}. Lam et al. conducted a study of five Microsoft projects, revealing that 995 out of 3,871 distinct builds (i.e., about 26\%) are affected by flaky tests~\cite{lam2019root}. This shows that even large organizations with rigorous processes face the challenge of flakiness in their test suites. 

Flaky tests can arise from multiple causes, including concurrency issues, test order dependencies, and platform dependencies~\cite{gruber2022survey,parry2021survey}. For example, Gruber et al.’s survey shows that concurrency issues are a major cause of flakiness in classical software~\cite{gruber2022survey}. Developers often express dissatisfaction with flaky tests occurring frequently, with some experiencing these inconsistencies on a daily basis~\cite{parry2021survey}. The unpredictable outcomes of flaky tests lead to resource-intensive debugging sessions and frequent reruns, which complicate the testing and debugging process. A variety of methods and tools exist to detect and mitigate flaky tests in classical software engineering. These include empirical studies like~\cite{luo2014empirical, gruber2021empirical}, flakiness detection tools~\cite{lam2019idflakies, silva2020shake, bell2018deflaker, alshammari2021flakeflagger, verdecchia2021know}, categorization tools~\cite{akli2023flakycat}, and mitigation strategies~\cite{barbosa2022test, habchi2022qualitative}.

As we transition into the era of quantum computing, quantum software has developed dramatically in terms of both numbers and complexity~\cite{de2022software}, which requires researchers and practitioners to design, develop, and maintain quantum software in a more standard and automated approach, just like what has happened in classical software engineering. Quantum mechanics (such as superposition and entanglement) also introduce new challenges in quantum software engineering, particularly in testing and debugging quantum software~\cite{zhao2020quantum}. For example, superposition allows quantum bits (qubits) to exist in multiple states simultaneously, which can give different outputs across identical test executions~\cite{valiev2005quantum}. Similarly, entanglement leads to a correlation between qubits that can vary based on measurement, increasing the likelihood of inconsistent test results~\cite{horodecki2009quantum}. 

Quantum flakiness detection faces new challenges. First, the inherent randomness of quantum software increases complexity in quantum testing and debugging, while most classical software does not have an indeterministic nature (except for probabilistic programs). Second, quantum-specific frameworks, such as Qiskit and NetKet, introduce unique syntactic and semantic patterns in Python, which are not explored in classical flakiness detection techniques. Third, compared to empirical studies in classical flakiness, quantum flakiness lacks empirical findings due to the limited number of quantum software projects and immaturity of quantum software engineering. 

We initiate a study on quantum flaky tests~\cite{zhang2023identifying} and identify the relationship between quantum indeterminacy and flakiness. Compared to classical flaky tests, which are usually caused by concurrency issues from testing, we identify the most common root cause for quantum flakiness is the ``randomness'' from both testing and the programs under testing~\cite{zhang2023identifying}. In~\cite{zhang2023identifying}, we search for 10 keywords related to flakiness, such as ``flaky'' and ``flakiness'', in issue reports (IRs) and pull requests (PRs) of 14 quantum software repositories on GitHub and find 46 unique flaky tests from 12 repositories. We also classify those quantum flaky tests into eight common causes, along with seven categories of fixes. This work lays the groundwork and provides valuable initial insights into the prevalence and nature of flakiness in quantum software. To the best of our knowledge, this is the only available quantum flaky test dataset. However, our previous study heavily relies on keyword-based methods to detect flaky tests, which may miss true cases of flakiness due to their dependence on specific vocabulary. The manual identification is also time- and resource-intensive, which makes it impractical for larger datasets and broader applications.

To overcome these limitations, we propose an automated quantum flakiness detection tool leveraging feature-based methods using machine learning (ML) approaches rather than keyword-matching to capture flakiness in quantum software. We train and evaluate multiple ML methods, i.e., eXtreme Gradient Boosting (XGB)~\cite{chen2016xgboost}, Decision Tree (DT)~\cite{quinlan1986induction},  Random Forest (RF)~\cite{breiman2001random}, $K$-Nearest Neighbor (KNN)~\cite{guo2003knn}, and Support Vector Machine (SVM)~\cite{cortes1995support}, in different performance metrics, such as F1 score and Matthews Correlation Coefficient (MCC). Our results indicate that the XGB and DT models outperform the other models in terms of F1 score and MCC, with feature-based techniques improving the reliability and efficiency of detecting flaky tests in quantum systems. Our long-term goals are to create strategies to detect different types of quantum flakiness and automatically recommend effective fixes.

Our \textbf{contributions} are twofold. First, we propose an automated method for detecting quantum flaky tests using ML techniques. Second, we extend the existing quantum flakiness dataset by introducing non-flaky tests, and we use this extended dataset to train and test our methods in both balanced and imbalanced approaches (to simulate more realistic real-world scenarios).

The rest of the paper is structured as follows. Section~\ref{sec:data} introduces our data preprocessing steps. Section~\ref{sec:model} introduces the models that we implement and the techniques adopted. Section~\ref{sec:training} explains the model training and evaluation processes. Section~\ref{sec:settings} illustrates the experimental settings. Section~\ref{sec:results} compares the results of all the model implementations and highlights the best-performing model. %
Section~\ref{sec:related-works} provides a literature review of related work in classical software flaky test detection and classification. Section~\ref{sec:threats} describes the threats to validity in our study. Section~\ref{sec:conclusion} concludes this paper, followed by the data availability in Section~\ref{sec:extra}.

\section{Data Preprocessing}\label{sec:data}

In this section, we outline the approaches implemented in this study to preprocess the data.

\subsection{Data Collection}

Our dataset is built upon our previous empirical study~\cite{zhang2023identifying}. They identify 46 unique flaky tests from 12 quantum software repositories using a keyword search of 10 terms related to flakiness: ``flaky'', ``flakiness'', ``flakey'', ``occasion'', ``occasional'', ``intermit'', ``fragile'', ``non-deterministic'', ``nondeterministic'', and ``brittle''. The artifacts are available on Zenodo, and the repository contains all IRs and PRs related to those 46 quantum flaky tests.\footnote{\url{https://zenodo.org/records/7888639}}

We are interested in the source code related to quantum flakiness. Thus, we screen all IRs and PRs in the existing dataset. First, a total of 99 files are extracted from those 46 flaky tests. Second, as we focus on Python files, we omit code files written in Q\# or unrelated to flakiness. Also, certain types of flaky files, particularly those related to environment setup, introduce noise to our dataset, hence decreasing the performance across all models. These environment-related flaky files are often caused by external factors, such as hardware variability, network instability, or dependency issues, leading to inconsistent test results unrelated to the code logic. After removing these files, we get a total of 45 unique Python files related to quantum flakiness from 6 repositories, namely \code{qiskit},\footnote{\url{https://github.com/Qiskit/qiskit}} \code{qiskit-ibm-provider},\footnote{\url{https://github.com/Qiskit/qiskit-ibm-provider}} \code{qiskit-ibmq-provider},\footnote{\url{https://github.com/Qiskit/qiskit-ibmq-provider}} \code{qiskit-ibm-runtime},\footnote{\url{https://github.com/Qiskit/qiskit-ibm-runtime}} \code{qiskit-nature},\footnote{\url{https://github.com/qiskit-community/qiskit-nature}} and \code{netket}.\footnote{\url{https://github.com/netket/netket}} Note that some of the quantum repositories have been restructured on GitHub, e.g., \code{qiskit-terra}, formerly the most active quantum repository with the maximum number of flaky tests, has been merged in \code{qiskit}. Thus, the repository names here are different than the names in~\cite{zhang2023identifying}. 

The artifacts from our previous study do not include any non-flaky tests~\cite{zhang2023identifying}. However, our study considers both flaky and non-flaky data for classification purposes. We collect non-flaky data in three steps. First, non-flaky tests are extracted by randomly sampling closed IRs and PRs from the \code{qiskit} repository, which is the most active quantum project in our study. Second, if a sampled IR or PR matches any of the 10 flaky keywords defined in~\cite{zhang2023identifying}, it is eliminated to ensure there are no flaky tests. We collect 100 IRs and PRs from which 257 non-flaky Python files are extracted. Third, the collected files are further analyzed to remove any file that contained environmental setup code, resulting in a final count of 243 non-flaky Python files for the study.

The extended dataset allows us to train models on both a balanced ratio (45 flaky and 45 non-flaky files, 1:1) and an imbalanced ratio (45 flaky and 243 non-flaky files, 1:5) to assess the performance of ML-based quantum flakiness detection techniques in different environments.

\subsection{Vectorization}

In order to convert the textual Python code files to a numeric format, we perform vectorization. The Python files are vectorized using the Bag-of-Words method~\cite{qader2019overview}, which compiles an unordered list of words from all the files and converts it into a document-term matrix, where each row represents an individual Python file, each column corresponds to the unique words in these files, and each cell marks the word occurrences. To improve the identification of code-specific keywords, we retain all the stop words, including terms such as ``if'', ``else'', and ``then'' during vectorization, as these words can provide important syntactical information in code files. 

\section{Model Implementation}\label{sec:model}

\begin{figure*}[htp!]
    \centering
    \includegraphics[width=0.95\linewidth]{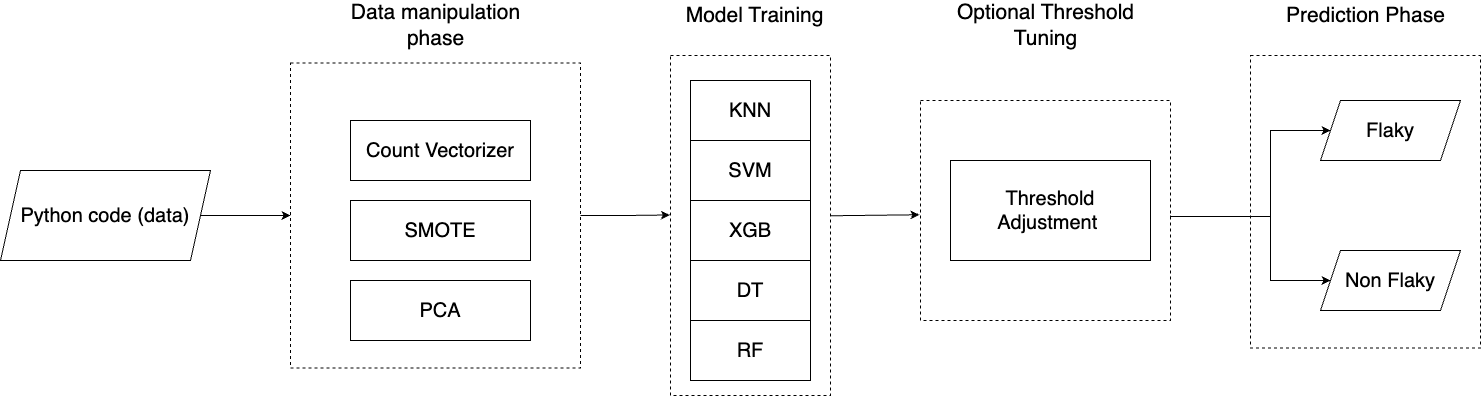}
    \caption{A high-level overview of our employed pipeline for flaky detection.}
    \label{fig:approach_pipelin}
\end{figure*}

As mentioned in Section~\ref{sec:Intro}, we employ five ML models in this study---XGB, DT, RF, KNN, and SVM. The selection of these models is based on our counterparts in classical flaky test detection tools~\cite{verdecchia2021know, alshammari2021flakeflagger}. For all models, we fine-tune their hyperparameters and find the best ones to optimize the results. In this section, we will explain the techniques used to implement those models.  %
Figure~\ref{fig:approach_pipelin} depicts an overview of the pipeline.

\subsection{Dimensionality Reduction}

In this research, the document-term matrix achieved after vectorization contains numerous features due to the diversity of tokens in the input code files. Thus, we apply Principal Component Analysis (PCA)~\cite{karamizadeh2013overview}---a dimensionality reduction technique that reduces the dimensions of the data while preserving the most important information from the original data and also eliminating feature redundancy. PCA is applied to the document-term matrix to reduce the large feature space while retaining features with higher variance to retain unique information in the reduced dataset. 

The maximum number of features is limited by the size of the dataset. In our five-fold cross-validation (CV) setting, the training set comprises 80 percent of the total dataset. Thus, in our experiment, the maximum number of PCA components is 230 out of 288 (the amount of observations in our dataset). We adopt grid search to determine the optimal number of principal components. 

However, we find that applying PCA to tree-based models, i.e., XGB, DT, and RF, decreases their performances even after fine-tuning. Thus, we choose not to apply PCA for the tree-based models because these models tend to handle high-dimensional data efficiently~\cite{biau2016random}.

\subsection{Handling Data Imbalance with SMOTE} \label{sec:SMOTE_approach}

We have seen the imbalance existing in quantum flaky tests~\cite{zhang2023identifying, zhang2024automated, sivaloganathan2024automating}. In order to handle the imbalanced ratio of 45 flaky to 243 non-flaky files, we employ Synthetic Minority Over-sampling Technique (SMOTE)~\cite{chawla2002smote}, a data augmentation method used to balance the data. SMOTE generates synthetic samples of the minority class (flaky tests in this case) by creating new data points. SMOTE identifies a number of nearest neighbors for a data point (we choose the default value of five) and generates a new data point between the original point and the randomly selected neighbor from the set. This process is repeated until the flaky data points match the number of non-flaky data points.

\subsection{Adjusting Decision Boundaries with Threshold Tuning} \label{sec:Thr_approach}

Besides SMOTE, we also employ threshold tuning with the imbalanced dataset for performance evaluation. In most cases, flaky classes get incorrectly labeled as non-flaky, leading to more false positives. Threshold tuning adjusts the decision boundary or sensitivity for classification to improve the balance between the precision and recall scores. In this study, we vary the threshold values ranging from 0.1 to 0.9 to find the optimal threshold for the performance metrics (the default setting is 0.5 for binary classifiers). %

\section{Model Training and Evaluation}\label{sec:training}

In this section, we provide a detailed explanation of the model training and evaluation processes.

\subsection{Model Training} 

Each model is trained through a five-fold stratified CV. We opt for stratified CV to keep the distribution of the classes uniform across folds. %

\textbf{Balanced dataset.} We first assess the performance of the models on a balanced dataset, i.e., 45 flaky and 45 non-flaky files. The data is vectorized by counting the occurrences of each unique word in all the input Python files. The textual Python code gets converted to a document-term matrix. For KNN and SVM, the vectorized training data is first passed through PCA, and the reduced data is then used to train the ML classifier, which classifies the test as flaky or non-flaky. The model is trained on a parameter grid consisting of various values of PCA components, specifying the count of reduced features and hyperparameters specific to each model. We use grid search to fine-tune the hyperparameters for each model, optimizing them based on the F1 score. The model is then tested on the unseen data, and the whole process is repeated until all five folds are processed. 

For XGB, DT, and RF, PCA is omitted to preserve the original data. Instead, the input data is vectorized, passed to the classifiers, and trained using a wide range of hyperparameters for each model in the five-fold CV.
 
\textbf{Imbalanced dataset.} We also investigate the performance of the vanilla models on the imbalanced (1:5) ratio, i.e., 45 flaky and 243 non-flaky files. In this setup, the initial training replicated the approach used in the balanced scenario, providing baseline performance metrics under the imbalanced condition. 

We adopt the imbalance dataset to test the models in a simulated real-world situation.\footnote{Note that the flaky test ratio varies on different projects; thus, our ratio here (i.e., 1:5) may not necessarily represent real-world cases. It rather represents an estimate to test the robustness of the models.} Imbalanced datasets can cause significant challenges (e.g., biased decision threshold and overfitting to the dominant class) for our classifiers, leading to suboptimal performance.  Our study utilizes two approaches for handling the imbalance: SMOTE augmentation and threshold tuning. Note that we apply SMOTE before PCA if PCA is in place. In order to identify the best technique for classification, this study implements the models in three different settings/configurations: 1) applying SMOTE, 2) applying threshold tuning, and 3) applying a hybrid approach, including both SMOTE and threshold tuning.

While SMOTE is applied during the training part to augment the data, thresholds are tuned during testing to make predictions. We aim to explore the benefits of training the models with both approaches to evaluate their effects on the model performances.

\subsection{Model Evaluation}\label{Model Evaluation}

We evaluate the performance of our methods through a five-fold stratified CV, and the average performance metrics (with standard deviations) are reported. We opt for accuracy, precision, recall, F1 score, and MCC metrics. The accuracy measures the overall proportion of correct predictions of both flaky and non-flaky tests. Precision measures the accuracy of flaky predictions; a higher precision score means fewer non-flaky tests are incorrectly classified as flaky. Recall measures the model’s ability to detect actual flaky tests; a higher recall means most flaky tests are detected correctly. The F1 score balances precision and recall to give accuracy for both flaky and non-flaky predictions. MCC evaluates the overall model performance by considering all possible outcomes.

\section {Experimental Settings}\label{sec:settings}

The hyperparameters of the ML algorithms are tuned to aim for the best F-1 score. Since SMOTE synthetically modifies the data by generating additional samples for the minority class, the data distribution and structure alter. As a result, the optimal hyperparameters for each model with SMOTE are different than the hyperparameters for vanilla models. Thus, we apply grid search in both the vanilla and SMOTE implementations.

\begin{itemize}

    \item \textbf{XGB} is configured with a learning rate of 0.5 to balance convergence speed and accuracy, a maximum tree depth of 5 to prevent overfitting, and 100 estimators for optimized performance.
    For models that utilize SMOTE, the maximum depth increases to 3 to balance convergence, with a learning rate of 0.3 and 200 estimators for optimized performance. 
    
    \item \textbf{DT} classifier utilizes the `entropy' criterion for measuring information gain during splits, a maximum tree depth of 10 to prevent overfitting. The model requires a minimum of 2 samples in each leaf node and at least 10 samples to split an internal node. 
    For models utilizing SMOTE, the criterion switches to `Gini', with the minimum samples per leaf node set to 2.
    
    \item \textbf{RF} classifier 
    configured with 200 estimators (trees) to enhance prediction stability and accuracy. It uses the `entropy' criterion for evaluating splits, a maximum tree depth of 10 to prevent over-fitting, and requires a minimum of 2 samples per leaf node and at least 5 samples to split an internal node.
    For models utilizing SMOTE, the number of estimators switches to 100, maximin tree depth to 10, and minimum sample splits to 5. 
    
    \item \textbf{KNN} uses the `Euclidean' metric for distance calculation, with 3 neighbors and distance-based weighting for predictions. Dimensionality reduction is applied using PCA with 150 components, ensuring efficiency and accuracy for high-dimensional data.   
    For models utilizing SMOTE, the number of neighbors increases to 7, the metric remains `Euclidean', and the number of PCA components shifts to 200, ensuring improved performance and compatibility with the adjusted dataset.
    
    \item \textbf{SVM} is configured with a regularization parameter of 0.01 to prevent overfitting and 220 components for dimensionality reduction using PCA. These settings ensure simplicity and effectiveness for high-dimensional datasets.
    For models that utilize SMOTE, the number of PCA components decreased to 180. 

\end{itemize}

\section {Experimental Results}\label{sec:results}

In this section, we will show the results of all five ML models on both balanced and imbalanced datasets. For imbalanced datasets, we will evaluate their performance in 1) vanilla, 2) SMOTE, 3) threshold tuning, and 4) hybrid implementations, respectively. 

\begin{table*}[thp!]
\centering
\caption{Performance metrics on balanced dataset. The values are reported in the average of the five folds and the standard deviation.}
\resizebox{\textwidth}{!}{%
\begin{tabular}{lccccc}
\toprule
\textbf{Model} & \textbf{Accuracy} & \textbf{Precision} & \textbf{Recall} & \textbf{F1} & \textbf{MCC} \\
\midrule
XGB & \textbf{0.933} ($\pm$ 0.042) & 0.924 ($\pm$ 0.069) & \textbf{0.956} ($\pm$ 0.089) & \textbf{0.934} ($\pm$ 0.043) & \textbf{0.877} ($\pm$ 0.075) \\
DT & 0.889 ($\pm$ 0.092) & \textbf{0.938} ($\pm$ 0.123) & 0.867 ($\pm$ 0.163) & 0.883 ($\pm$ 0.103) & 0.805 ($\pm$ 0.156) \\
RF & 0.889 ($\pm$ 0.050) & 0.878 ($\pm$ 0.071) & 0.911 ($\pm$ 0.083) & 0.891 ($\pm$ 0.050) & 0.786 ($\pm$ 0.100) \\

KNN           & 0.744 ($\pm$ 0.056) & 0.872 ($\pm$ 0.108) & 0.600 ($\pm$ 0.151) & 0.690 ($\pm$ 0.106) & 0.525 ($\pm$ 0.099) \\
SVM           & 0.833 ($\pm$ 0.086) & 0.858 ($\pm$ 0.096) & 0.800 ($\pm$ 0.130) & 0.824 ($\pm$ 0.101) & 0.673 ($\pm$ 0.171) \\
\bottomrule
\end{tabular}%
}
\label{tab:equal_model_performance}
\end{table*}

\subsection{Vanilla Models}

Our initial experiments involved training the vanilla models on a balanced dataset with a 1:1 ratio of flaky to non-flaky tests. Table~\ref{tab:equal_model_performance} summarizes the performance metrics for each model in the averages of the five folds and the standard deviations. Overall, XGB outperforms the other models in almost all five metrics (except for precision). More specifically, the XGB achieves the highest F1 score of 0.934 and the highest MCC of 0.877, respectively. %

For the imbalanced dataset, the performance of the vanilla models can be found in Table~\ref{tab:model_metrics_imbalance}. As a comparison of the vanilla models in Tables~\ref{tab:equal_model_performance} and~\ref{tab:model_metrics_imbalance}. In terms of the F1 score, the DT and RF models' performance drops slightly, but the performance of XGB, KNN, and SVM drops significantly from a balanced to an imbalanced scenario. In terms of MCC, the performance of DT and RF actually improves, the performance of KNN and SVM stays about the same, and the performance of XGB drops significantly (from 0.877 to 0.441) from a balanced to an imbalanced scenario. 

Among all vanilla models in Table~\ref{tab:model_metrics_imbalance}, we observe that the DT model performs the best in terms of both F1 score (0.877) and MCC (0.864).

\begin{table*}[htp!]
\caption{Performance Metrics on Imbalanced Dataset under Different Methods. The best results among all models are highlighted with underscores.}
\centering
\small
\resizebox{\textwidth}{!}{
\begin{tabular}{@{} l l p{2.2cm} p{2.2cm} p{2.2cm} p{2.2cm} p{2.2cm} @{}}
\toprule
\textbf{Method} & \textbf{Model} & \textbf{Accuracy} & \textbf{Precision} & \textbf{Recall} & \textbf{F1} & \textbf{MCC} \\
\midrule

\multirow{5}{*}{\textbf{Vanilla}}
&XGB & \underline{\textbf{0.980}} ($\pm$ 0.040) & 0.778 ($\pm$ 0.122) & 0.859 ($\pm$ 0.055) & 0.850 ($\pm$ 0.055) & 0.441 ($\pm$ 0.091) \\
& DT & 0.962 ($\pm$ 0.025) & 0.913 ($\pm$ 0.121) & \textbf{0.867} ($\pm$ 0.129) & \textbf{0.877} ($\pm$ 0.079) & \textbf{0.864} ($\pm$ 0.087) \\
& RF & 0.961 ($\pm$ 0.020) & \textbf{0.946} ($\pm$ 0.065) & 0.800 ($\pm$ 0.083) & 0.866 ($\pm$ 0.082) & 0.849 ($\pm$ 0.083) \\
& KNN           & 0.892 ($\pm$ 0.013) & 0.920 ($\pm$ 0.098) & 0.356 ($\pm$ 0.109) & 0.497 ($\pm$ 0.110) & 0.522 ($\pm$ 0.073) \\
& SVM           & 0.920 ($\pm$ 0.024) & 0.845 ($\pm$ 0.094) & 0.622 ($\pm$ 0.194) & 0.691 ($\pm$ 0.128) & 0.672 ($\pm$ 0.112) \\

\midrule

\multirow{5}{*}{\textbf{SMOTE only}}
& XGB & \textbf{0.969} ($\pm$ 0.023) & \underline{\textbf{0.978}} ($\pm$ 0.044) & 0.822 ($\pm$ 0.151) & \textbf{0.884} ($\pm$ 0.096) & \textbf{0.877} ($\pm$ 0.094) \\
& DT & 0.955 ($\pm$ 0.042) & 0.920 ($\pm$ 0.160) & \textbf{0.844} ($\pm$ 0.206) & 0.850 ($\pm$ 0.144) & 0.845 ($\pm$ 0.140) \\
& RF & 0.944 ($\pm$ 0.013) & 0.824 ($\pm$ 0.048) & 0.822 ($\pm$ 0.054) & 0.822 ($\pm$ 0.054) & 0.790 ($\pm$ 0.049) \\
& KNN & 0.872 ($\pm$ 0.044) & 0.605 ($\pm$ 0.138) & 0.622 ($\pm$ 0.206) & 0.592 ($\pm$ 0.144) & 0.531 ($\pm$ 0.162) \\
& SVM           & 0.920 ($\pm$ 0.024) & 0.845 ($\pm$ 0.094) & 0.622 ($\pm$ 0.194) & 0.691 ($\pm$ 0.128) & 0.672 ($\pm$ 0.112) \\

\midrule

\multirow{5}{*}{\textbf{Threshold only}}
& XGB             & 0.962 ($\pm$ 0.013) & \textbf{0.964} ($\pm$ 0.073) & 0.800 ($\pm$ 0.130) & 0.863 ($\pm$ 0.056) & 0.854 ($\pm$ 0.055) \\
& DT & \textbf{0.965} ($\pm$ 0.027) & 0.938 ($\pm$ 0.137) & \textbf{0.866} ($\pm$ 0.144) & \underline{\textbf{0.886}} ($\pm$ 0.083) & \underline{\textbf{0.877}} ($\pm$ 0.089) \\
& RF & 0.961 ($\pm$ 0.023) & 0.946 ($\pm$ 0.073) & 0.800 ($\pm$ 0.092) & 0.866 ($\pm$ 0.082) & 0.849 ($\pm$ 0.093) \\
& KNN           & 0.875 ($\pm$ 0.031) & 0.650 ($\pm$ 0.152) & 0.556 ($\pm$ 0.136) & 0.579 ($\pm$ 0.094) & 0.521 ($\pm$ 0.100) \\
& SVM           & 0.920 ($\pm$ 0.015) & 0.756 ($\pm$ 0.050) & 0.733 ($\pm$ 0.169) & 0.733 ($\pm$ 0.082) & 0.694 ($\pm$ 0.086) \\
 
\midrule

\multirow{5}{*}{\textbf{Hybrid}}
& XGB & 0.\textbf{969} ($\pm$ 0.023) & \textbf{\underline{0.978}} ($\pm$ 0.044) & 0.822 ($\pm$ 0.151) & \textbf{0.884} ($\pm$ 0.096) & \textbf{0.877} ($\pm$ 0.094) \\
& DT & 0.956 ($\pm$ 0.035) & 0.898 ($\pm$ 0.155) & \underline{\textbf{0.889}} ($\pm$ 0.122) & 0.876 ($\pm$ 0.091) & 0.863 ($\pm$ 0.098) \\
& RF & 0.944 ($\pm$ 0.013) & 0.824 ($\pm$ 0.048) & 0.822 ($\pm$ 0.054) & 0.822 ($\pm$ 0.054) & 0.790 ($\pm$ 0.049) \\
& KNN & 0.899 ($\pm$ 0.033) & 0.713 ($\pm$ 0.084) & 0.578 ($\pm$ 0.215) & 0.623 ($\pm$ 0.156) & 0.580 ($\pm$ 0.161) \\
& SVM           & 0.937 ($\pm$ 0.014) & 0.820 ($\pm$ 0.092) & 0.800 ($\pm$ 0.163) & 0.792 ($\pm$ 0.067) & 0.768 ($\pm$ 0.069) \\

\midrule

\end{tabular}}
\label{tab:model_metrics_imbalance}
\end{table*}

\subsection{SMOTE Models}
To address the data imbalance issue, we apply the SMOTE method as described in Section~\ref{sec:SMOTE_approach}. In terms of F1 score and MCC, we observe a significant improvement from the vanilla version for XGB. For example, the value of MCC increases to 0.877 from 0.441. However, we do not observe significant changes by adopting SMOTE for the other four models. For example, both the F1 score and MCC of DT and RF decrease slightly, and the F1 score and MCC of SVM stay the same from the vanilla model to the SMOTE model.   %

To summarize, XGB and DT models still outperform RF, KNN, and SVM  models with SMOTE, while XGB outperforms all the other models in four metrics (except recall).

\subsection{Threshold Tuning Models}
As described in~\ref{sec:Thr_approach}, we tune the decision threshold of our models to adjust the classification sensitivity of the models without altering the dataset, and we choose the thresholds giving the best F1 scores. %
As demonstrated in Table~\ref{tab:model_metrics_imbalance}, threshold tuning also leads to enhancements in models' F1 scores compared with the vanilla models. For example, the F1 score has improved from 0.850 to 0.863 for XGB, from 0.877 to 0.886 for DT, from 0.497 to 0.579 for KNN, and from 0.691 to 0.733 for SVM. For RF, we observe no changes in the F1 score (the fine-tuned threshold is still 0.5 in this case). 

In general, DT provides the best performance in terms of F1 score and MCC, while XGB achieves the highest precision. 

This improvement suggests that adjusting the classification threshold allows the model to be more sensitive to the minority class, resulting in better detection of quantum flaky tests.

\subsection{Hybrid Models}
In the hybrid models (with both SMOTE and threshold tuning), all models provide stable performance compared to vanilla models in terms of F1 score and MCC, except for RF, where the value of MCC drops from 0.849 to 0.790. The biggest improvement that we observe is from XGB, and its MCC value increases from 0.441 to 0.877 (by 99\%). 

In general, XGB outperforms the other models in terms of F1 score and MCC, while DT achieves the best recall.   %

\subsection{Discussions}\label{sec:discussion}
The results of our experiments reveal several key insights. First, in all experiments, XGB and DT consistently outperform RF, KNN, and SVM in both balanced and imbalanced scenarios. The strong performance of these tree-based models aligns with previous findings of classical flaky test detection reported in~\cite{alshammari2021flakeflagger}.

Second, we observe that the performance of all vanilla models drops from the balanced dataset to the imbalanced dataset in terms of recall and F1 score. This shows that (for certain performance metrics) some ML models struggle with imbalanced and skewed datasets when no balancing technique is applied. For example, the MCC of XGB drops from 0.877 to 0.441 when the dataset becomes imbalanced. 

Third, the DT model achieves the best overall performance in the imbalanced dataset with the threshold tuning configuration, with an F1 score of 0.886 and an MCC of 0.877 (the standard deviation is 0.089, which is lower compared to the other two same scores). This demonstrates the capabilities of the XGB and DT models, especially DT, to detect flaky tests correctly, even in imbalanced scenarios with threshold adjustments. 

Last but not least, the results also highlight the importance of selecting appropriate techniques to address data imbalance. While DT and RF inherently handle data imbalanced effectively, XGB, KNN, and SVM benefit from the SMOTE and threshold tuning techniques. %

\section{Related Work}\label{sec:related-works}

The software quality becomes uncertain if test cases unpredictably change their outcomes (such as from pass to fail or the reverse) without any changes to the codebase, known as flaky tests. A typical approach to identify test flakiness is to re-run the tests to confirm if their outcomes are consistent. The iDFlakies~\cite{lam2019idflakies} identifies flaky tests by running tests in different randomized orders and tracking inconsistencies in test outcomes. When a test’s outcome changes between runs, iDFlakies flags it as potentially flaky. The framework also partially classifies the flaky tests by identifying patterns in test order dependencies, helping developers understand the flakiness causes. Shaker~\cite{silva2020shake} detects concurrency-related flaky tests by systematically manipulating thread schedules during test execution, a technique known as schedule perturbation. By altering the order of thread execution, Shaker increases the likelihood of exposing concurrency issues, such as race conditions, deadlocks, and timing dependencies, that might otherwise go unnoticed. However, the rerunning approach is costly; developers invest significant time and effort in diagnosing and investigating the root causes of flaky tests. Thus, the re-running approach is an inefficient and time-consuming method for managing test flakiness. 

To address the limitations of the re-running approach, many static methods are proposed, including ML-based approaches~\cite{bell2018deflaker, alshammari2021flakeflagger, verdecchia2021know}. DeFlaker~\cite{bell2018deflaker} was proposed to detect flaky tests by monitoring code coverage information during a single test run. It tracks which lines of code are covered by each test and identifies cases where a test fails without covering any new code compared to previous runs. This approach suggests that such failures are likely due to flakiness, as there are no relevant code changes that could explain the failure. The FlakeFlagger~\cite{alshammari2021flakeflagger} framework predicts the likelihood of a test being flaky by analyzing static features extracted from code and test history (past pass/fail history, frequency of recent changes, and frequency of test failures over time). The framework leverages ML binary classifiers trained on attributes associated with flakiness, enabling it to flag potentially flaky tests proactively without multiple executions. They experimented with several ML classifiers, including RFs, XGB, and SVM. Feature selection and hyperparameter tuning were performed to improve the model’s ability to generalize and reduce overfitting. Additionally, FLAST~\cite{verdecchia2021know} introduces a static analysis approach to predict test flakiness without reruns, leveraging features from neighboring code dependencies and characteristics. The model uses static features—like code complexity, dependency patterns, and the flakiness history of neighboring code—to predict if a test will be flaky. This neighbor-based analysis assumes that tests closely tied to flaky code are also likely to be flaky.

Several studies focus on detecting the root causes of flaky tests using ML approaches. For instance, FlakyCat~\cite{akli2023flakycat} leverages ML models to identify the categories of flaky tests. The framework uses supervised learning techniques, with a pre-defined taxonomy of flaky test types as labels. Their results demonstrated that ML-based categorization is a promising approach to categorizing flaky tests effectively. We avoid reviewing more papers in this regard, as it is out of the scope of the main goal of this study.

Although ML-based approaches can greatly reduce the overhead linked to dynamic methods, they still have certain limitations. Specifically, these models often struggle with imbalanced data in flaky test detection, as flaky instances are rare in real-world scenarios. The imbalance can cause ML models to favor stability in their predictions, limiting their ability to reliably detect instances of flakiness. In our study, we aim to reflect this imbalance issue in our studied dataset and investigate the effectiveness of integrating two existing methods—threshold tuning and SMOTE—within our proposed pipelines.

\section{Threats to validity}\label{sec:threats}

Validity threats are classified according to~\cite{wohlin2012experimentation,yin2009case}.

\textbf{Internal.} Flaky tests in Qiskit repositories range from 0.25\% to 1.09\%~\cite{zhang2023identifying}, highlighting a significant real-world imbalance. To partially reflect this issue in our studied dataset, we designed our dataset with non-flaky tests, outnumbering flaky tests by approximately a factor of five. To address this issue in our pipeline, we applied two complementary techniques, SMOTE and threshold tuning. While SMOTE generates synthetic examples of flaky tests to improve model training stability and reduce bias, threshold tuning allows us to adjust model sensitivity to better identify flaky tests within an imbalanced dataset. Our goal is to evaluate the effectiveness of these techniques in managing imbalance rather than precisely replicating real-world ratios---a goal we plan to address in future work. 

Other potential internal validity threats include the impact of selected hyperparameters in our ML algorithms and the accuracy of their performance measurements. To mitigate these risks, we performed a grid search to identify the optimal parameter configurations and employed stratified CV to ensure balanced and accurate evaluations. These steps, along with the application of rigorous, tailored validation strategies designed to address our specific research questions, help ensure the reliability and robustness of our findings.

\textbf{External.} Software engineering studies often face challenges related to real-world variability, and the issue of generalization can only be partially addressed~\cite{wieringa2015six}. 
The scope of our dataset is limited to a subset of quantum software repositories, specifically focusing on Python files from the \code{qiskit} and \code{netket} GitHub repository. Consequently, the results may not fully reflect the broader spectrum of quantum software projects.

\section{Conclusions and Future Work}\label{sec:conclusion}
In this paper, we proposed an ML-based classification method for detecting flaky tests in quantum programs. We extended the previous dataset and trained five widely used ML models, i.e., XGB, DT, RF, KNN, and SVM, to classify tests as flaky or non-flaky on both balanced and imbalanced scenarios. Our findings revealed that in the balanced dataset scenario, the tree-based models outperformed the non-tree-based models in all five performance metrics. More specifically, XGB achieves the highest performance in four metrics. In the imbalanced scenario, we also observed the best performance from XGB and DT, and the DT model (with the threshold tuning) provided the best F1 score and MCC. %

In the future, we aim to expand the dataset with more quantum flaky and non-flaky tests to reflect real-world scenarios. We also plan to explore other unsupervised ML techniques, such as large language models, to detect and classify quantum flaky tests and develop a more effective method to predicate quantum flakiness. We also aim to integrate our framework with continuous integration pipelines to automate the detection of flakiness in new source codes.

\section{Data Availability}\label{sec:extra}

The dataset, program code, and research materials are publicly available on Zenodo as an open-access project (accessible at \url{https://doi.org/10.5281/zenodo.14232772}). %

\bibliographystyle{IEEEtran}
\bibliography{references}

\end{document}